\documentclass[submission]{eptcs}
\providecommand{\event}{UITP 2012} 

\usepackage{amsmath}
\usepackage{amsfonts}
\usepackage{amssymb}
\usepackage{graphicx}
\usepackage{textcomp}
\usepackage{MnSymbol}

\newcommand{\tma}{\emph{Theorema}}
\newcommand{\tmaone}{\emph{Theorema 1.0}}
\newcommand{\tmatwo}{\emph{Theorema 2.0}}
\newcommand{\tc}{\emph{Theorema} commander}

\def\Quant#1#2{\begin{array}[t]{@{}c@{}}#1 \\[-1.5ex] \scriptstyle #2 \end{array}\;}
\def\ForAll#1{\Quant{\forall}{#1}}
\def\Let#1{\Quant{\text{let}}{#1}}
\def\QuantCond#1#2#3{\begin{array}[t]{@{}c@{}}#1 \\[-1.5ex] \scriptstyle #2 \\[-1.ex] \scriptstyle #3 \end{array}\;}
\def\ForAllCond#1#2{\QuantCond{\forall}{#1}{#2}}

\title{\tmatwo: A Graphical User Interface for a Mathematical Assistant System}
\author{Wolfgang Windsteiger\\
\institute{RISC, JKU Linz\\
4232 Hagenberg, Austria}\\
\email{Wolfgang.Windsteiger@risc.jku.at \qquad www.risc.jku.at/home/wwindste/}}
\def\titlerunning{\tmatwo: A Graphical User Interface for a Mathematical Assistant System}
\def\authorrunning{W.~Windsteiger}

\begin{document}

\maketitle

\begin{abstract}
{\tmatwo} stands for a re-design including a complete re-implementation of the {\tma} system, which was originally designed, developed, and implemented by Bruno Buchberger and his {\tma} group at RISC. In this paper, we present the first prototype of a graphical user interface (GUI) for the new system. It heavily relies on powerful interactive capabilities introduced in recent releases of the underlying Mathematica system, most importantly the possibility of having dynamic objects connected to interface elements like sliders, menus, check-boxes, radio-buttons and the like. All these features are fully integrated into the Mathematica programming environment and allow the implementation of a modern user interface.
\end{abstract}

\section{Introduction}
\label{sec:Intro}

The reasons for re-designing the {\tma} system are manifold. Some of them refer to the usability of the system from the users' point of view, others are related to the flexibility from the developers' point of view such as limitations in the combination of provers. Finally, using the system and observing other users working with the system over the time has shown several possibilities for improvements that cannot be easily realized without reconsideration of principal design decisions which have been made fifteen years ago. For these reasons, we decided to re-implement {\tma} with the aim to re-design those components, that have turned out to be the main hurdles, and reuse design ideas\footnote{What we reuse are \emph{the ideas} only. We do not reuse any existing Mathematica code from {\tmaone}, because we want to change some internal datastructures and cleanup the code base on that occasion as well.} that have proved successful and useful.

A crucial decision in every software development project is, of course, the choice of the development platform. {\tmaone} was based on Mathematica \cite{Mma3}, one of the leading computer algebra systems developed by Wolfram Research, mainly for two reasons: firstly, because of its convenient programming language offering the powerful pattern matching mechanism, which is extremely well-suited for the implementation of logical inference rules, and secondly for the nice notebook user interface. The availability of a huge library of symbolic mathematical algorithms does not harm, but it is not and it never was the crucial factor in favor of Mathematica. Maybe the only drawback is the commercial setting, a {\tma} user needs to purchase a license of Mathematica in order to be able to run {\tma}. But there are additional arguments on the pro-side, such as platform independence, i.e Mathematica programs run without any modifications on essentially all available operating system platforms (Linux, OS X, and Windows), the 
powerful development group at Wolfram Research that keeps Mathematica being always an up-to-date platform growing into various directions, and the huge group of Mathematica users also outside the classical theorem proving community. We did have minor compatibility issues when new releases of Mathematica arrived, but in retrospect we probably had fewer problems over more than fifteen years than we would have had on comparable platforms. We therefore decided to stay with Mathematica also for the implementation of {\tmatwo}. In this paper, we concentrate on the novelties related to the user interface exclusively.

{\tmaone} \cite{RISC2487,RISC2330,RISC2674,RISC2737} has been widely acknowledged as a system with one of the nicer user interfaces. However, we could observe that outsiders or beginners still had a very hard time to successfully use the {\tma} system. This was true for entering formulae correctly as well as for proving theorems or performing computations. The principal user interface to {\tma} is given by the Mathematica notebook front-end, giving a small advantage to Mathematica users as they are familiar with the main interaction patterns offered by the notebook interface. While the 2D-syntax for mathematical formulae available since Mathematica~3 is nice to read, a wrongly entered 2D-structure has always been a common source of errors. More than that, the user-interaction paradigm in {\tmaone} was the standard `command-evaluate' known from Mathematica, meaning that every action in {\tmaone} was triggered by the evaluation of a certain {\tma} command implemented as a 
Mathematica program. As an example, giving a definition meant evaluation of a Definition[\ldots]-command, stating a theorem meant evaluation of a 
Theorem[\ldots]-command, proving a theorem meant evaluation of a Prove[\ldots]-command, and performing a computation meant evaluation of a Compute[\ldots]-command. For the new {\tmatwo} system, we envisage a more `point-and-click'-like interface as one is used to from modern software tools like a mail user agent or office software. 

The main target user group for {\tma} are mathematicians, who want to engage in formalization of mathematics or who just want to have some computer-support in their proofs. The system should be a tool helping to grasp the nature of proving, thus, students of mathematics or computer science are typical users as well as teachers at universities or high schools. For the latter groups in particular, nice two-dimensional input and output of formulae in an appearance like typeset or handwritten mathematics is an important feature. On the other hand, the unambiguous parsing of mathematical notation is non-trivial already in 1D, supporting 2D-notations introduces some additional difficulties.

{\tma} is a multi-method system, i.e.~it offers many different proving methods specialized for the proof task to be carried out. The main focus lies on a resulting proof that comes as close as possible to a proof done by a well-educated mathematician. This results in a multitude of methods, each of them having a multitude of options to fine-tune the behavior of the provers. This is on the one hand powerful and gives many possibilities for system insiders, who know all the tricks and all the options including the effect they will have in a particular example. For newcomers, on the other hand, the right choice of an appropriate method and a clever choice of option settings is often an insurmountable hurdle. The user interface in {\tmatwo} makes these selections easier for the user. Furthermore, the user has the possibility to extend the system by self-defined reasoning rules and strategies.

Finally, the integration of proving, computing, and solving \emph{in one system} will stay a major focus also in {\tmatwo}. Compared to 
{\tmaone}, the separation between {\tma} and the underlying Mathematica system is even stricter, but the integration of Mathematica's computational facilities into the {\tma} language has been improved.

{\tmatwo} is currently under development. The components described in this paper are all implemented and the screenshots provided show a running and working system, it is not the sketch of a design. However, the interface presented here is incomplete and it will grow with new demands. From the experience with Mathematica's GUI components gathered up to now we are confident that all requirements for a modern interface to a mathematical assistant system can easily be fulfilled based on that platform. Some of the features described in this paper rely or depend on their implementation in Mathematica. This requires a certain knowledge of the principles of Mathematica's programming language and user front-end in order to understand all details given below.
The rest of the paper is structured as follows: the first section describes the new features in recent releases of Mathematica that form the basis for new developments in {\tmatwo}, in the second section we introduce the new {\tma} user interface, and in the conclusion we give a perspective for future developments.

\section{New in Recent Versions of Mathematica}

In this section, we describe some of the new developments in recent Mathematica releases that were crucial in the development of {\tmatwo}.

\subsection{Mathematica Dynamic Objects}

Earlier versions of Mathematica offered the so-called \emph{GUIKit} extension, which was based on Java and used MathLink for communication between Mathematica and the generated GUI. We used GUIKit earlier for the development of an educational front-end for {\tma} \cite{RISC3067}, but the resulting GUI was cumbersome to program, unstable, and slow in responding to user interaction. As of Mathematica version~6, and then reliably in version~7, the concept of \emph{dynamic expressions} was introduced into the Mathematica programming language and fully integrated into the notebook front-end. Dynamic expressions form the basis for interactive system components, thus, they are \emph{the} elementary ingredient for the new {\tmatwo} GUI.

In short, every Mathematica expression can be turned into a dynamic object by wrapping it into \texttt{Dynamic}. As the most basic example, \texttt{Dynamic[}$expr$\texttt{]} produces an object in the Mathematica front-end that displays as $expr$ and automatically updates as soon as the value $expr$ changes. In addition, typical interface elements such as sliders, menus, check-boxes, radio-buttons, and the like are available. Every such element can then be connected to a program variable, such that user interactions (e.g.~clicking a check-box) are reflected in the values of the respective variables. The set of available GUI objects is very rich and there is a wide variety of options and auxiliary functions in order to influence their behavior and interactions. These features allow the construction of arbitrarily complicated dynamic interfaces and seem to constitute a perfect platform for the 
implementation of an interface to the {\tma} system. A big advantage of this approach is that the entire interface programming can be done inside the Mathematica environment, which in particular brings us a uniform interface on all platforms from Linux over Mac until Windows for free.

\subsection{Cascading Stylesheets}

Stylesheets are a means for defining the appearance of Mathematica notebook documents very similar to how stylesheets work in HTML or word processing programs. The mere existence of a stylesheet mechanism for Mathematica notebooks is not new, but what is new since version~6 is that stylesheets are cascading, i.e.~stylesheets may depend on each other and may inherit properties from their underlying styles just like CSS in HTML. This of course facilitates the design of different styles for different purposes without useless duplication of code. The more important news is that stylesheets can now, in addition to influencing the appearance of a cell in a notebook, also influence the \emph{behavior} of a cell. This is a feature that we always desired since the beginning of {\tma}: an action in Mathematica is always connected in some way to the evaluation of a cell in a notebook, and we wanted to have different evaluation behavior depending on whether we want to e.g. prove something, do a computation, enter a 
formula, or execute an algorithm. Using a stylesheet, we can now define computation-cells or formula-cells, and the stylesheet defines commands for their pre-processing, evaluation, and post-processing.

\section{The {\tma} Interface}
\label{sec:TmaInterface}

\begin{figure}[htb]
  \centering
  \includegraphics[width=\textwidth]{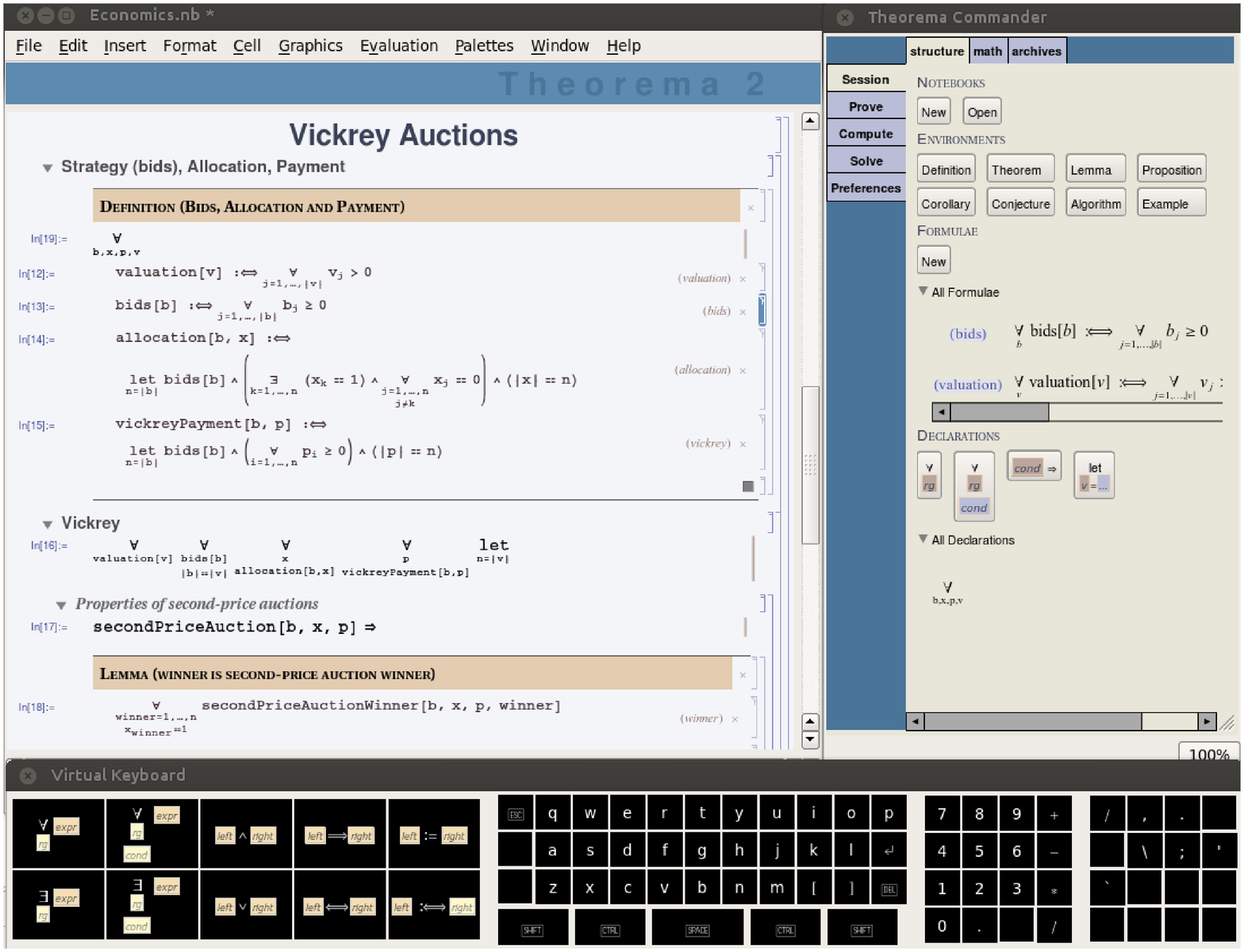}
  \caption{The {\tmatwo} GUI}
  \label{fig:GUI}
\end{figure}

The Mathematica notebook front-end is the primary user interface for {\tma}. ``Working in {\tma}'' consists of \emph{activities} that themselves require certain \emph{actions} to be performed. As an example, a typical activity would be ``to prove a theorem'', which requires actions such as ``selecting a proof goal'', ``composing the knowledge base'', ``choosing the inference rules and a proof strategy'', etc. The central new component in {\tmatwo} is the {\tma} \emph{commander}; it is the GUI component that guides and supports all activities and actions. Of course, most activities work on mathematical formulae in one or the other way. Formulae appear as definitions, theorems or similar \emph{environments} and are just written into Mathematica/{\tma} notebook documents that use one of the {\tma} stylesheets. As {\tma} \emph{session} we refer to the collection of all formulae passed to the system up to a certain moment. Composing and manipulating the session is just another activity and therefore 
supported from the {\tc}. The second new interface component in {\tmatwo} is the \emph{virtual keyboard}; its task is to facilitate the input of math expressions, in particular 2D-input. Figure~\ref{fig:GUI} shows a screen shot of {\tmatwo} with a {\tma}-styled notebook\footnote{The actual  mathematics written in the notebook is irrelevant for this paper, but for the curious it is part of a formalization of auction theory, an important application of mathematics in economy. This is joint work with M.~Kerber, C.~Lange, and C.~Rowat at the University of Birmingham \cite{RISC4610}.} top-left, the {\tc} to its right, and the virtual keyboard underneath. Of course, all these features are just add-ons to the standard Mathematica interface, thus, support for notebook formatting, inputting special characters, text styling, and the like through notebook menus, palettes, and/or keyboard shortcuts need not be implemented from the scratch.

\subsection{Organizing a {\tma} Session}
\label{sec:Env}

When working in {\tma} one composes one or more {\tma}-styled Mathematica notebooks, which have all the capabilities of normal Mathematica notebooks plus the possibility to process expressions in {\tma} language inside so-called \emph{formula cells}. This means that {\tma} expressions are embedded in a full-fledged document format for mathematical writing. Mathematica notebooks consist of hierarchically arranged cells, whose nesting is visualized with cell brackets on the right margin of the notebook. Figure~\ref{fig:GUI} shows a notebook using one of the {\tma}-specific stylesheets responsible for the notebook's appearance and behavior. Note in particular that, due to this stylesheet, each environment forms a group for its own.

{\tma} formula cells contain mathematical expressions in {\tma} syntax with an additional label. If no label is given by the user, a numerical label, which is unique within the notebook, is automatically assigned. User-supplied labels need not be unique, but the system issues a warning to the user. As soon as the formula is passed to the system through Mathematica's standard Shift-Enter, the formula is stored in an internal datastructure that carries a \emph{unique key} for each formula in addition to the formula itself and its label. This key consists of the absolute pathname of the notebook file in which it was given, and the unique cell-ID within that notebook, which is provided by the Mathematica front-end. In formula display, we typically use the label, but when actually referring to a formula in the interface, we use the unique formula key. As we will explain later, the user never sees nor needs the concrete formula key explicitly.

In mathematical practice, universal quantification of formulae and conditioning is often done on a global level. As an example take definitions, which often start with a phrase like ``Let $n\in\mathbb{N}$. We then define \ldots'', which in effect expresses a universal quantifier for $n$ plus the condition $n\in\mathbb{N}$ for all notions introduced in the current definition. For this purpose, we provide \emph{global declarations}, which may either contain one or several ``orphaned'' universal quantifiers (each containing a variable and an optional condition, but missing the formula, to which they refer) or an ``orphaned'' implication (missing its right hand side), or an abbreviation indicated by a ``let''. The idea is that the scope of such a declaration ranges to the end of the environment in which it appears. In the example in Figure~\ref{fig:GUI}, this is used in \textsc{Definition (Bids, Allocation and Payment)} with a universal quantifier for $b,x,p$, and $v$ valid for all formulae inside that 
definition. When passing to the system e.g.~the formula written in the notebook as
$$\text{bids}[b]:\Longleftrightarrow \ForAll{j=1,\ldots,|b|}b_j\geq 0$$
it actually results in
$$\ForAll{b}\text{bids}[b]:\Longleftrightarrow \ForAll{j=1,\ldots ,|b|}b_j\geq 0$$
being stored in the {\tma} session. For the user's convenience, the {\tc} always shows all formulae currently available in the section labeled `All Formulae' as shown in Figure~\ref{fig:GUI}. There one can also see, that quantifiers are of course only put for those variables that actually appear free in the formula. The cell grouping defined in the stylesheet ensures that a definition gets its own cell group that limits the scope of the quantifier.

We generalized this idea so that a global declaration can be put anywhere in a notebook, and its scope ranges similar to the situation described above from its position to the end of the nearest enclosing cell group. In Figure~\ref{fig:GUI}, this is used twice:
\begin{enumerate}
  \item There is a big 
  \[\ForAll{\text{valuation}[v]}\ForAllCond{\text{bids}[b]}{|b|\Leftrightline|v|}\ForAllCond{x}{\text{allocation}[b,x]}\ForAllCond{p}{\text{vickreyPayment}[b,p]}\Let{n=|v|}\]
  at the beginning of Section~`Vickrey'. This means that, without further mentioning, all free occurrences of $v,b,x$, and $p$ will be universally quantified with the respective additional conditions in the entire section including all its subsections. Furthermore, wherever we write $n$ it is just an abbreviation for $|v|$.
  \item There is a `$\text{secondPriceAuction}[b,x,p]\Rightarrow$' in Subsection~`Properties of second-price auctions', so that this condition on $b,x$, and $p$ affects only the rest of this subsection.
\end{enumerate}
At the moment of passing a formula to the system, all declarations valid at this position are silently applied and the actual formula in the {\tma} session has all intended quantifiers and conditions attached to it just as if they were written explicitly with the formula. Thus, the Lemma compactly written as 
\[
  \ForAllCond{\text{winner}=1,\ldots,n}{x_{\text{winner}}\Leftrightline 1}
  \text{secondPriceAuctionWinner}[b,x,p,\text{winner}]
\]
in the notebook in Figure~\ref{fig:GUI} actually states
\begin{multline*}
    \ForAll{\text{valuation}[v]}\ForAllCond{\text{bids}[b]}{|b|\Leftrightline|v|}\ForAllCond{x}{\text{allocation}[b,x]}
    \ForAllCond{p}{\text{vickreyPayment}[b,p]}\\
  \text{secondPriceAuction}[b,x,p]\Rightarrow \ForAllCond{\text{winner}=1,\ldots,|v|}{x_{\text{winner}}\Leftrightline1}
  \text{secondPriceAuctionWinner}[b,x,p,\text{winner}].
\end{multline*}

This is quite convenient and comes very close to how mathematicians are used to write down things. In essence, the effect of global declarations is similar to what can be achieved with contexts or locales in Isabelle \cite{Kammueller99locales:a}. For bigger documents, however, one might lose the overview on which declarations are valid at a certain point in a notebook. The {\tc} gives valuable assistance in this situation: the section labeled `All Declarations' always shows all declarations valid at the current cursor position in the selected notebook. In Figure~\ref{fig:GUI}, the selection is at the cell containing the definition of $\text{bids}[b]$ within the Definition-environment, and correspondingly the {\tc} displays the $\ForAll{b,x,p,v}$ valid there. 

\subsection{The {\tma} Commander}\label{sec:TmaCom}

Figure~\ref{fig:GUI} top-right shows the {\tc}, the main GUI component in {\tmatwo}. It is a two-level tabview structured according to activities on the first level and the corresponding actions for each activity on the second level. The first-level activity-tabs reside on the left margin of the {\tc}. Currently, the supported activities are `Session', `Prove', `Compute', `Solve', and `Preferences', but as the system develops, this list may increase. For each of these activities, the respective actions can be accessed on the top margin of the {\tc}. Moving through them from left to right corresponds to a wizard guiding the user through the respective activity. Proving is presumably the most interesting activity and we will therefore elaborate it in more detail in the next paragraph. The remaining parts of the {\tc} are of similar fashion, we will only mention some highlights in the concluding paragraph of this section.

\paragraph{The `Prove'-activity}

\begin{figure}[htb]
  \centering
  \includegraphics[width=0.45\textwidth]{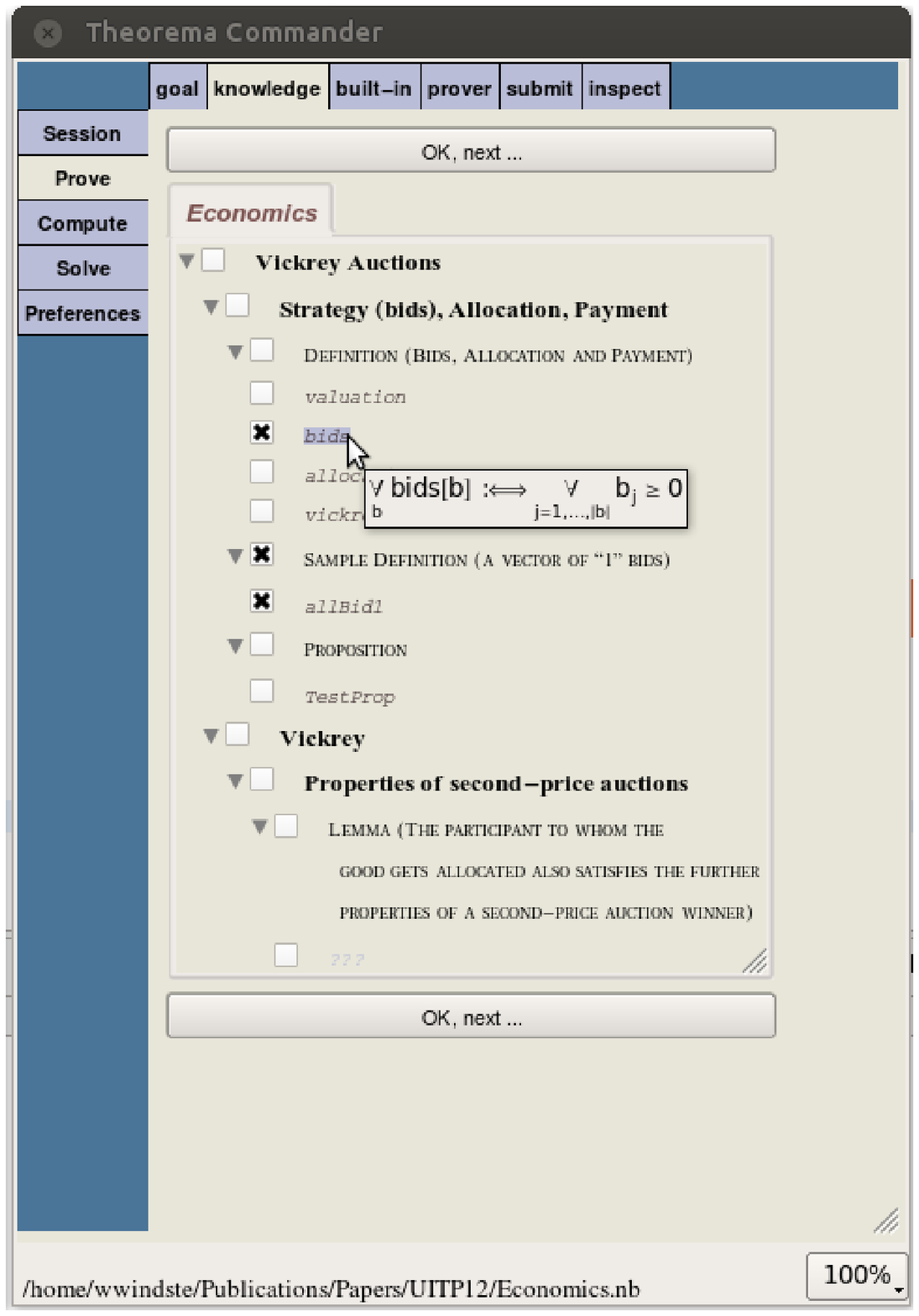} \hspace*{10mm}
  \includegraphics[width=0.45\textwidth]{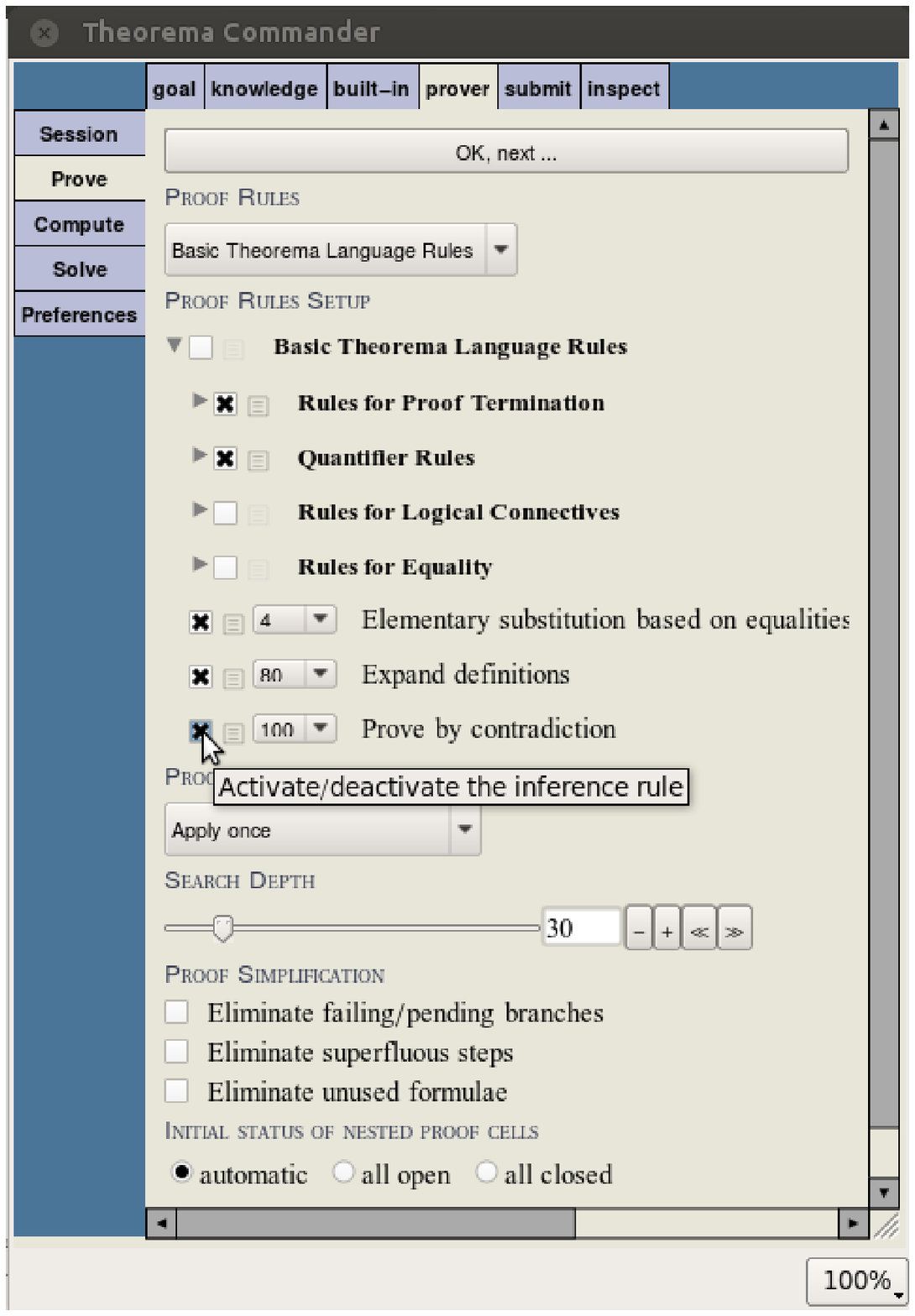}
  \caption{The `Prove'-activity: the knowledge browser (left) and the prover configuration (right).}
  \label{fig:ProveI}
\end{figure}

The `Prove'-activity consists of actions `goal', `knowledge', `built-in', `prover', `submit', and `inspect', see the screenshots in Figure~\ref{fig:ProveI}. These action correspond to the individual steps when proving a theorem in {\tma}: it requires the specification of the proof goal, the specification of the knowledge available in the proof, setting up built-in knowledge, and selecting/configuring the desired prover to be used. After submitting the proof problem to {\tma}, the system will show a successful or failing proof, which the user can then inspect.

Defining the proof goal is as simple as just selecting a cell containing the formula to be proved in an open notebook with the mouse. The selected formula is then shown in the `goal'-tab, and it changes with every mouse selection. The only action required here is to confirm the choice by pressing a button in the `goal'-tab. From this moment on, the proof goal is fixed until the next confirmation, whatever the mouse selects.
  
Then the user needs to compose the knowledge base to be used in the proof, see Figure~\ref{fig:ProveI} (left). The \emph{knowledge browser} displays a tab for each open notebook or loaded knowledge archive\footnote{\emph{Archives} are another new development in {\tmatwo}. An archive gives the possibility to store the formulae from a notebook efficiently in an external file, such that they can be loaded quickly into a {\tma} session. Since this is not a user-interface issue, we do not go into further details here.}. In each tab, a hierarchical overview of the file/archive content is displayed, showing only the section structure, environments, and formula labels. Simply moving the mouse cursor over the label opens a tooltip displaying the whole formula, clicking the label jumps to the respective position in the corresponding notebook/archive. Each entry in the browser has a check-box attached to its left responsible for toggling the selection of the respective unit. In this way, individual formulae, environments, sections, up to 
entire notebooks can be selected or deselected with just one mouse-click. The formulae chosen in this way constitute the knowledge base for the next proof. The formula label displayed in the browser is only syntactic sugar, the check-box is connected to the unique key of each formula in the {\tma} session, see Section~\ref{sec:Env}.

The next action within the `Prove'-activity is the selection of built-in computational knowledge\footnote{With \emph{built-in knowledge} we refer to knowledge built into the {\tma} language semantics. As an example, `$+$' is by default an uninterpreted operator. Using some built-in knowledge one can link `$+$' to the addition of numbers available in the {\tma} language. This is a feature inherited from {\tmaone}.}. The \emph{built-in browser} works like the knowledge browser described above. Instead of section grouping we have (not necessarily disjoint) thematic groups of built-ins like sets, arithmetic, or logic. Built-in knowledge is applied in proving in order to simplify formulae by computation on finite objects, e.g.~computations with numbers or finite sets. We do not go into further details.

After having composed the relevant built-in knowledge, the user needs to select the prover. A \emph{prover} in {\tmatwo} consists of a (possibly nested) list of inference rules accompanied with a proof strategy. Accordingly, the `prover'-action shows menus for choosing the inference rules and the strategy, respectively, together with short info panels explaining the current choice as depicted in Figure~\ref{fig:ProveI} (right). The `prove'-action displays an \emph{inference rule browser} corresponding to the selected rule list. Its functionality is like that of the knowledge browser described above, only that it is using the nesting structure of the inference rule list for setting up the hierarchy, which gives the possibility to activate/deactivate entire groups with only one click. Using the inference rule browser the user can efficiently deactivate individual (groups of) inference rules, e.g.~for influencing whether an implication will be proved directly or via contraposition. In addition to the checkbox for 
activation and deactivation, the interface allows to decide whether the respective proof step should be explained in the final proof or not. This is an easy way to set the granularity of the resulting natural language explanation of the proof. Moreover, the priority of each rule in the underlying proof search can be adjusted through a popup-menu. Again, all interface elements are explained by tooltips as soon as the mouse moves over them.

\begin{figure}[htb]
  \centering  
  \includegraphics[width=0.95\textwidth]{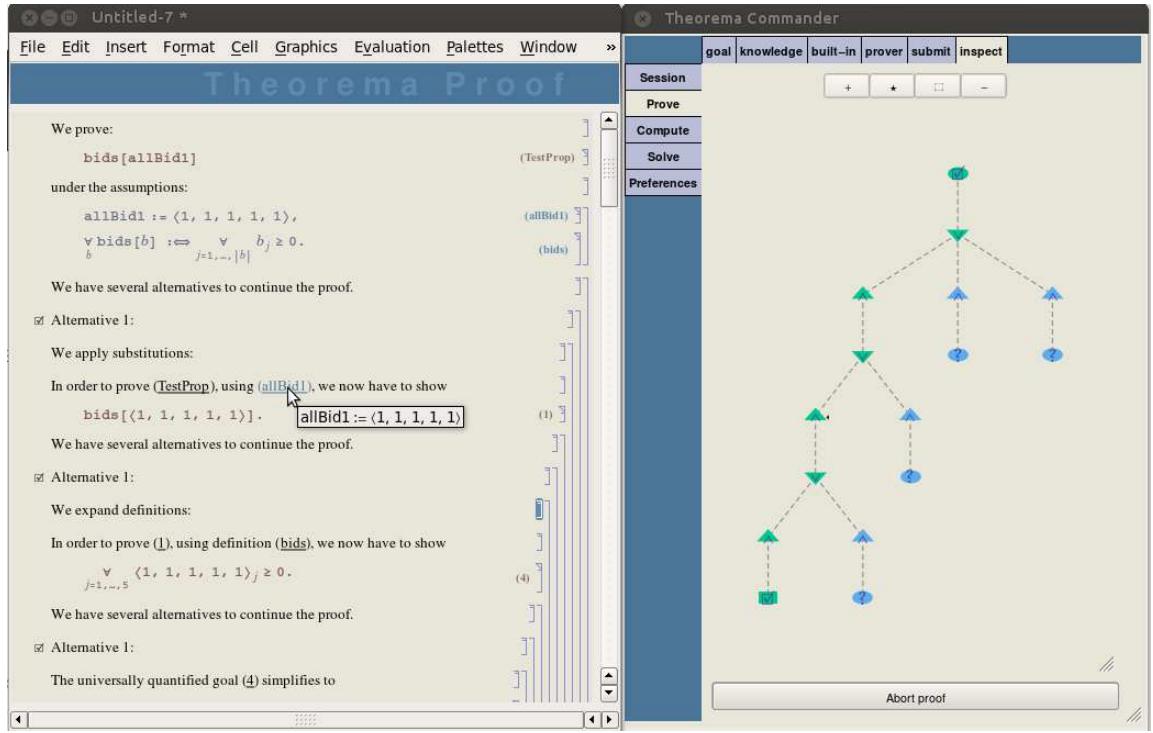}
  \caption{The `Prove'-activity: a generated proof (left) and the corresponding `inspect'-action (right).}
  \label{fig:ProveII}
\end{figure}

Once the prover is configured, the proof task is ready to be submitted. The `submit'-action collects all settings from the previous actions, in particular the chosen goal and knowledge base, and displays them for a final check. Hitting the `Prove'-button submits all data to the {\tma} kernel and automatically proceeds to the `inspect'-action. Figure~\ref{fig:ProveII} (right) displays the corresponding proof tree as it develops during proof generation. The nodes in the proof tree differ in shape, color, and content depending on node type and status. As soon as the proof is finished, some proof information is written back into that notebook, in which the proof goal has been stated. In addition to an indicator of proof success or failure and a summary of settings used at the time of proof generation, this information contains two important buttons:
\begin{enumerate}
  \item A button to display the proof in natural language in a separate window as shown in Figure~\ref{fig:ProveII} (left). This feature is in essence the same as we had it in {\tmaone}~\cite{RISC2737}. The `inspect'-tab in the {\tc} and the proof display are connected in both directions: clicking a node in the proof tree jumps to the respective text blocks in the proof display describing the corresponding proof step; clicking a cell in the proof display marks the corresponding tree node with a small black triangle. In combination this offers a nice possibility to navigate through a proof. As one can see from Figure~\ref{fig:ProveII}, all formula labels used in the natural proof presentation use tooltips to show the full formula, to which they refer.
  \item A button to restore all settings in the {\tc} to the values they had at the time of proof generation, which is a quick way to rerun a proof.
\end{enumerate}

\paragraph{Other activities}

The `Session'-activity consists of structuring formulae into definitions, theorems, etc., arranging global declarations (see Section~\ref{sec:Env}), inspecting the session, inputting formulae, and the development and maintenance of knowledge archives. In the `Compute'-activity, a user sets up the expression to be computed and selects the knowledge base and the built-in knowledge to be used in the computation (using knowledge- and built-in browsers as described for proving above). Knowledge selections for proving are independent from those used for computations.

In the `Preferences'-activity we collect everything regarding system setup, such as e.g.~the preferred language. The entire GUI is language independent in the sense that no single English string (for GUI labels, button labels, explanations, tooltips, etc.) is hardcoded in its implementation, but all strings are constants, whose definitions are collected in several language-setup files. For effective language translation it is important that users have access to the language-setup files so that every user has the possibility to translate the system into her language and that new languages can be integrated with minimal effort. The {\tmatwo} architecture is such that the language selection menu in the `Preferences' will offer the choice among all languages, for which a setup file is present (in a certain directory). This has the effect that, for the translation into a new language, only the English files have to be copied and renamed, and the English texts need to be translated. Without any further action, the 
new language can be selected from the menu, and voil{\`a} the GUI runs in the new language.

Some other aspects of internationalization are already solved by Mathematica, e.g. the availability of language dependent special characters, unicode, country-specific number formating, etc., others will be considered in future work, e.g. placement of buttons and the ``logical direction'' of action-wizards for languages written from right to left. In particular for educational purposes that we envisage for {\tmatwo}, internationalization is of utmost importance.

An important detail that makes all this possible is the decision to license {\tmatwo} under GPL\footnote{The system will be available from GITHUB by mid-July 2013, the {\tma} homepage \url{http://www.risc.jku.at/research/theorema/software/}  will provide more information from then on.}. This gives all users access not only the language-setup files but to the entire source code. An attractive perspective for user contribution to the system could then also be the development of new inference rules or proof strategies. These are just Mathematica programs, and there is a rich library of {\tma} programs that is ready for use in the implementation of inference rules and strategies.

\subsection{The Virtual Keyboard}

The last component to be described briefly is the \emph{virtual keyboard}, see the screenshot in Figure~\ref{fig:GUI}. Although much of the typical input can be given through buttons and palettes, it turns out that still the keyboard is the most efficient way to enter expressions, at least once a user is a little familiar with the system. Therefore, the {\tma}-stylesheets define keyboard shortcuts for the most frequently used {\tma} expressions. In the absence of a physical keyboard---e.g.~when working on a tablet computer or on an interactive whiteboard in an educational context---we provide the virtual keyboard, which is an arrangement of buttons imitating a physical keyboard. It consists of a character block for the usual letters and a numeric keypad (numpad) for digits and common arithmetic operators like on common keyboards. As a generalization of the numpad, we provide a \emph{sympad} (to the far right) and an \emph{expad} (to the left) for common mathematical symbols and expressions, respectively. 
Using modifier keys like 
Shift, Mod, Ctrl and more, every key on the board can be equipped with many different meanings depending on the setting of the modifiers. We believe that the virtual keyboard is a very powerful input component for mathematical expressions, which will prove useful even in the presence of a physical keyboard, where the buttons react to mouse-clicks.  

\section{Conclusion}

Some of the features are implemented currently as `proof of concept' and need to be completed in the near future to get a system that can be used for case studies. As an example, the {\tma} language syntax, from parsing via formatted output to computational semantics, is only implemented for a fraction of what we already had in {\tmaone}. Due to the fact that the already implemented parts are the most complicated ones and that we paid a lot of attention to a generic programming style, we are optimistic that progress can be made quickly in that direction.

The bigger part of the work to be done is the re-implementation of all provers that we already had in {\tmaone}. What we already have now is the generic proof search procedure and the mechanism of inference rule lists and strategies with their interplay. Two sample strategies, one that models more or less the strategy used in {\tmaone} and another one that does a more fine-grained branching on alternative inference rules being applicable, are already available, but no report on their performance can be given at this stage. The big effort is now to provide all the inference rules for standard predicate logic including all the extensions that the {\tma} language supports. As soon as this is completed we can engage in case studies trying out the system in some real-world theory formalization and in education, for which we plan a hybrid interactive-automatic proof strategy to be available. Towards university education in mathematics and logic, we see a big potential for an interactive proof-assistant based on the new user interface, in particular the proof-tree navigation presented in Section~\ref{sec:TmaCom}.

\bibliography{uitp}

\end{document}